\documentclass[preprint,showpacs,preprintnumbers,amsmath,amssymb]{revtex4}
\usepackage{graphicx,epsfig,dcolumn,bm,epic,eepic,float}% Include figure files
\usepackage{amsmath}
\usepackage{latexsym}
\usepackage{color}
\usepackage{cancel}
\usepackage{makeidx,shortvrb,latexsym}
\begin{document}
\unitlength 1 cm
\newcommand{\nn}{\nonumber}
\newcommand{\vk}{\vec k}
\newcommand{\vp}{\vec p}
\newcommand{\vq}{\vec q}
\newcommand{\vkp}{\vec {k'}}
\newcommand{\vpp}{\vec {p'}}
\newcommand{\vqp}{\vec {q'}}
\newcommand{\bk}{{\bf k}}
\newcommand{\bp}{{\bf p}}
\newcommand{\bq}{{\bf q}}
\newcommand{\br}{{\bf r}}
\newcommand{\bR}{{\bf R}}
\newcommand{\up}{\uparrow}
\newcommand{\down}{\downarrow}
\newcommand{\fns}{\footnotesize}
\newcommand{\ns}{\normalsize}
\newcommand{\cdag}{c^{\dagger}}

\title {$KMR$ $k_t$-factorization procedure for the description of the $LHCb$ forward hadron-hadron $Z^0$ production at $\sqrt{s}=13\;TeV$}
\author{$M. \; Modarres$}
\altaffiliation {Corresponding author, Email: mmodares@ut.ac.ir,
Tel:+98-21-61118645, Fax:+98-21-88004781.}
\author{$M.R. \; Masouminia$}
\author{$R. \; Aminzadeh\;Nik$}
\affiliation {Department of Physics, University of $Tehran$,
1439955961, $Tehran$, Iran.}

\begin{abstract}
Quit recently, two sets of new experimental data from the $LHCb$ and
the $CMS$ collaborations have been published, concerning the
production of the $Z^0$ vector boson in  hadron-hadron collisions
with the center-of-mass energy $E_{CM}= \sqrt{s}=13\;TeV$. On the
other hand, in our recent work, we have conducted a set of $NLO$
calculations for the production of the electroweak gauge vector
bosons, utilizing the unintegrated parton distribution functions
($UPDF$) in the frameworks of $Kimber$-$Martin$-$Ryskin$ ($KMR$) or
$Martin$-$Ryskin$-$Watt$ ($MRW$) and the $k_t$-factorization
formalism, concluding that the results of the $KMR$ scheme are
arguably better in describing the existing experimental data, coming
from $D0$, $CDF$, $CMS$ and $ATLAS$ collaborations. In the present
work, we intend to follow the same $NLO$ formalism and calculate the
rate of the production of the $Z^0$ vector boson, utilizing the
$UPDF$ of $KMR$ within the dynamics of the recent data. It will be
shown that our results are in good agreement with the new
measurements of the $LHCb$ and the $CMS$ collaborations.
\end{abstract}
 \maketitle

\section{Introduction}
\label{sec:I}

Traditionally, the production of the electroweak gauge vector bosons
is considered as a benchmark for understanding the dynamics of the
strong and the electroweak interactions in the Standard Model.  It
is also an important test to assess the validity of collider data.
Many collaborations have reported numerous sets of measurements,
probing different events in variant dynamical regions, in direct or
indirect relation with such processes, to count a few see the
references \cite{ATLAS-2012, ATLAS-2015, CMS-2011, ATLAS2016,
CMS2015, LHCb-2012, LHCb-2013, LHCb-20151, LHCb-20152, LHCb-20153}.
Among the most recent of these reports are the measurements of the
production of $Z^0$ bosons at the $LHCb$ and $CMS$ collaborations,
for proton-proton collisions at the $LHC$ for $\sqrt{s} = 13TeV$,
with different kinematical regions \cite{LHCb,CMS}. The $LHCb$ data
are in the forward pseudorapidity region ($2<|\eta|<4.5$) while the
$CMS$ measurements are in th central domain ($0<|\eta|<2.4$).

In our previous work \cite{NLO-W/Z}, we have successfully utilized
the transverse momentum dependent ($TMD$) unintegrated parton
distribution functions ($UPDF$) of the $k_t$-factorization (the
references \cite{KMR,MRW}), namely the $Kimber$-$Martin$-$Ryskin$
($KMR$) and $Martin$-$Ryskin$-$Watt$ ($MRW$) formalisms in the
leading order ($LO$) and the next-to-leading order ($NLO$) to
calculate the inclusive production of the $W^{\pm}$ and the $Z^0$
gauge vector bosons, in the proton-proton and the proton-antiproton
inelastic collisions
\begin{equation}
    P_1 + P_2 \to W^{\pm}/Z^0 + X.
    \label{eq1}
\end{equation}
In order to increase the precision of the calculations, we have used
a complete set of $2 \to 3$ $NLO$ partonic sub-processes, i.e.
\begin{eqnarray}
    g^{*}(\textbf{k}_1) + g^{*}(\textbf{k}_2) \to V(\textbf{p}) + q(\textbf{p}_1) + \bar{q}'(\textbf{p}_2), \nonumber \\
    g^{*}(\textbf{k}_1) + q^{*}(\textbf{k}_2) \to V(\textbf{p}) + g(\textbf{p}_1) + q'(\textbf{p}_2), \nonumber \\
    q^{*}(\textbf{k}_1) + \bar{q}^{'*}(\textbf{k}_2) \to V(\textbf{p}) + g(\textbf{p}_1) + g(\textbf{p}_2),
    \label{eq2}
\end{eqnarray}
where $V$ represents the produced gauge vector boson. $\textbf{k}_i$
and $\textbf{p}_i$, $i=1,2$ are the 4-momenta of the incoming and
the out-going partons. The results underwent comprehensive and
rather lengthy comparisons and it was concluded that the
calculations in the $KMR$ formalism are more successful in
describing the existing experimental data (with the center-of-mass
energies of $1.8$ and $8$ TeV) from the $D0$, $CDF$, $ATLAS$ and
$CMS$ collaborations
\cite{ATLAS2016,CMS2015,CDF96,CDF2000,D095,D098,D02000-1,D02000-2,D02001}.
The success of the $KMR$ scheme (despite being of the $LO$ and
suffering from some misalignment with its theory of origin, i.e. the
$Dokshitzer$-$Gribov$-$Lipatov$-$Altarelli$-$Parisi$ ($DGLAP$)
evolution equations, \cite{DGLAP1,DGLAP2,DGLAP3,DGLAP4}) can be
traced back to the particular physical constraints that rule its
kinematics. To find extensive discussions regarding the structure
and the applications of the $UPDF$ of $k_t$-factorization, the
reader may refer to the references \cite{Modarres1, Modarres2,
Modarres3, Modarres4, Modarres5, Modarres6, Modarres7, Modarres8}.

Meanwhile, arriving the new data from the $LHCb$ and $CMS$
collaborations, the references \cite{LHCb,CMS}, gives rise to the
necessity of repeating our calculations at the $E_{CM}=13\;TeV$.
This is in part due to the very interesting rapidity domain of the
$LHCb$ measurements, since in the forward rapidity sector
($2<|\eta_f|<4.5$), one can effectively probe very small values of
the Bjorken variable $x$ ($x$ being  the fraction of the
longitudinal momentum of the parent hadron, carried by the parton at
the top of the partonic evolution ladder), where the gluonic
distributions dominate and hence the transverse momentum dependency
of the particles involving in the partonic sub-processes becomes
important.

In the present work, we intend to calculate the transverse momentum
and the rapidity distributions of the cross-section of production of
the $Z^0$ boson using our $NLO$ level diagrams (from the reference
\cite{NLO-W/Z}) and the $UPDF$ of the $KMR$ formalism. The $UPDF$
will be prepared using the $PDF$ of $MMHT2014-LO$, \cite{MMHT}. In
the following section, the reader will be presented with a brief
introduction to the $NLO \otimes LO$ framework (i.e. $NLO$ $QCD$
matrix elements and $LO$ $UPDF$) that is utilized to perform these
computations. The section \ref{sec:II} also includes the main
description of the $KMR$ formalism in the $k_t$-factorization
procedure. Finally, the section \ref{sec:III}  is devoted to
results, discussions and a thoroughgoing conclusion.
\section{$NLO \otimes LO$ framework, $KMR$ $UPDF$ and numerical analysis}
\label{sec:II}
Generally speaking, the total cross-section for an
inelastic collision between two hadrons ($\sigma_{Hadron-Hadron}$)
can be expressed as a sum over all possible partonic cross-sections
in every possible momentum configuration:
\begin{eqnarray}
    \sigma_{Hadron-Hadron} =&& \sum_{a_1,a_2=q,g} \int_0^1 {dx_1 \over x_1} \int_0^1 {dx_2 \over x_2}
    \int_{0}^{\infty} {dk^2_{1,t} \over k^2_{1,t}} \int_{0}^{\infty} {dk^2_{2,t} \over k^2_{2,t}}
    f_{a_1}(x_1,k^2_{1,t},\mu_1^2) f_{a_2}(x_2,k^2_{2,t},\mu_2^2)
    \nonumber \\
    \times &&
    \hat{\sigma}_{a_1 a_2}(x_1,k^2_{1,t},\mu_1^2;x_2,k^2_{2,t},\mu_2^2).
    \label{eq3}
\end{eqnarray}
In the equation (\ref{eq3}), $x_i$ and $k_{i,t}$ respectfully
represent the longitudinal fraction and the transverse momentum of
the parton $i$, while $f_{a_i}(x_i,k^2_{i,t},\mu_i^2)$ are the
density functions of the $i^{th}$ parton. The second scale, $\mu_i$,
are the ultra-violet cutoffs related to the virtuality of the
exchanged particle (or particles) during the inelastic scattering. $
\hat{\sigma}_{a_1 a_2}$ are the partonic cross-sections of the given
particles. For the production of the $Z^0$ boson, the equation
(\ref{eq3}) comes down to (for a detailed description see the
reference \cite{NLO-W/Z})
    $$
    \sigma (P+\bar{P} \rightarrow Z^0 + X) = \sum_{a_i,b_i = q,g} \int
    {dk_{a_1,t}^2 \over k_{a_1,t}^2} \; {dk_{a_2,t}^2 \over k_{a_2,t}^2} \;
    dp_{b_1,t}^2 \; dp_{b_2,t}^2 \; dy_1 \; dy_2 \; dy_{W/Z}
    \; \times
    $$
    $$
    {d\varphi_{a_1} \over 2\pi} \; {d\varphi_{a_2} \over 2\pi} \; {d\varphi_{b_1} \over 2\pi} \; {d\varphi_{b_2} \over 2\pi}
    \times
    $$
    \begin{equation}
    {|\mathcal{M}(a_1+a_2 \rightarrow Z^0+b_1+b_2)|^2 \over 256 \pi^3 (x_1 x_2 s)^2} \;
    f_{a_1}(x_1,k_{a_1,t}^2,\mu^2) \; f_{a_2}(x_2,k_{a_2,t}^2,\mu^2).
    \label{eq4}
    \end{equation}
$y_i$ are the rapidities of the produced particles (since $y_i\simeq
\eta_i$ in the infinite momentum frame, i.e. $p_i^2 \gg m_i^2$).
$\varphi_i$ are the azimuthal angles of the incoming and the
out-going partons at the partonic cross-sections. $|\mathcal{M}|^2$
represent the matrix elements of the partonic sub-processes in the
given configurations. The reader can find a number of comprehensive
discussions over the means and the methods of deriving analytical
prescriptions of these quantities in the references
\cite{NLO-W/Z,LIP1,LIP2,LIP3,Deak1}. $s$ is the center of mass
energy squared. Additionally, in the proton-proton center of mass
frame, one can utilize the following definitions for the kinematic
variables:
\begin{eqnarray}
    &&P_1 = {\sqrt{s} \over 2} (1,0,0,1), \;\;\; P_2 = {\sqrt{s} \over 2} (1,0,0,-1),
    \nonumber \\
    &&\textbf{k}_i = x_i \textbf{P}_i + \textbf{k}_{i,\perp}, \;\;\; k_{i,\perp}^2 = -k_{i,t}^2,  \;\;\; i=1,2 \; .
    \label{eq5}
\end{eqnarray}
Defining the transverse mass of the produced particles,
$m_{i,t}=\sqrt{m^2_i + p_{i}^2}$, we can write,
\begin{eqnarray}
    x_1 &&= {1 \over \sqrt{s}} \left( m_{1,t} e^{+y_1} + m_{2,t} e^{+y_2} + m_{Z,t} e^{+y_{Z}} \right),
    \nonumber \\
    x_2 &&= {1 \over \sqrt{s}} \left( m_{1,t} e^{-y_1} + m_{2,t} e^{-y_2} + m_{Z,t} e^{-y_{Z}} \right).
    \label{eq6}
\end{eqnarray}

Furthermore, the density functions of the incoming partons,
$f_a(x,k^2_{t},\mu^2)$ (which represent the probability of finding a
parton at the semi-hard process of the partonic scattering, with the
longitudinal fraction $x$ of the parent hadron, the transverse
momentum $k_t$ and the hard-scale $\mu$) can be defined in the
framework of $k_t$-factorization, through the $KMR$ formalism:
\begin{equation}
    f_a(x,k_t^2,\mu^2) = T_a(k_t^2,\mu^2) \sum_{b=q,g} \left[ {\alpha_S(k_t^2) \over 2\pi}
    \int^{1-\Delta}_{x} dz P_{ab}^{(LO)}(z) {x \over z} b\left( {x \over z}, k_t^2 \right) \right],
    \label{eq7}
\end{equation}
The $Sudakov$ form factor, $T_a(k_t^2,\mu^2)$, factors over the
virtual contributions from the $LO$ $DGLAP$ equations, by defining a
virtual (loop) contributions as:
    \begin{equation}
    T_a(k_t^2,\mu^2) = exp \left( - \int_{k_t^2}^{\mu^2} {\alpha_S(k^2) \over 2\pi}
    {dk^{2} \over k^2} \sum_{b=q,g} \int^{1-\Delta}_{0} dz' P_{ab}^{(LO)}(z') \right), \label{eq8}
    \end{equation}
with $T_a(\mu^2,\mu^2) = 1$. $\alpha_S$ is the $LO$ $QCD$ running
coupling constant, $P_{ab}^{(LO)}(z)$ are the so-called splitting
functions in the $LO$, parameterizing the probability of finding a
parton with the longitudinal momentum fraction $x$ to be emitted
form a parent parton with the fraction $x'$, while $z=x/x'$, see the
references \cite{MRW,PNLO}. The infrared cutoff parameter, $\Delta$,
is a visualization of the angular ordering constrtaint ($AOC$), as a
consequanse of the color coherence effect of successive gluonic
emittions \cite{KMR1}, defined as $\Delta = {k_t /( \mu + k_t)}$.
Limiting the upper boundary on $z$ integration by $\Delta$, excludes
$z=1$ form the integral equation and automatically prevents facing
the soft gluon singularities, \cite{NLO-W/Z}. Additionally, the
$b(x,k_t^2)$ are the single-scaled parton distribution functions
($PDF$), i.e. the solutions of the $LO$ $DGLAP$ evolution equation.
The required $PDF$ for solving the equation (\ref{eq7}) are provided
in the form of phenomenological libraries, e.g. the $MMHT2014$
libraries, the reference \cite{MMHT}, where the calculation of the
single-scaled functions have been carried out using the deep
inelastic scattering data on the $F_2$ structure function of the
proton.

Now, one can carry out the numerical calculation of the equation
(\ref{eq4}) using the $\mathtt{VEGAS}$ algorithm in the  Monte-Carlo
integration, \cite{VEGAS}. To do this, we have chosen the hard-scale
of the $UPDF$ as:
    $$
    \mu = (m_{W/Z}^2 + p_{W/Z,t}^2 )^{1 \over 2},
    $$
and set the upper bound on the transverse momentum integrations of
the equation (\ref{eq4}) to be $k_{i,max} = p_{i,max} = 4
\mu_{max}$, with
    $$
    \mu_{max} = (m_{W/Z}^2 + p_{t,max}^2 )^{1 \over 2}.
    $$
One can easily confirm that since the $UPDF$ of $KMR$ quickly vanish
in the $k_t \gg \mu$ domain,  further domain have no contribution
into our results. Also we limit the rapidity integrations to
$[-8,8]$, since $0 \leq x \leq 1$ and according to the equation
(\ref{eq6}), further domain has no contribution into our results.
The  choice of above hard scale is reasonable for the production of
the Z bosons, as has been discussed in the reference \cite{Deak1}.

Finally, we choose
\begin{equation}
    f_{a_i}(x_i,k_{a_i,t}^2<\mu_0^2,\mu^2) = {k_{a_i,t}^2 \over \mu_0^2} a_i(x_i,\mu_0^2) T_{a_i}(\mu_0^2,\mu^2),
    \label{eq9}
\end{equation}
to define the density of the incoming partons in the
non-perturbative region, i.e. $k_t<\mu_0$ with $\mu_0=1\;GeV$. This
appears to be a natural choice, since (see the references
\cite{NLO-W/Z,WattWZ})
    $$
    \lim_{k_{a_i,t}^2 \rightarrow 0} f_{a_i}(x_i,k_{a_i,t}^2,\mu^2) \sim k_{a_i,t}^2.
    $$

\section{Results, Discussions and Conclusions}
\label{sec:III} Using the theory and the notions of the previous
sections, one can calculate the production rate of the $Z^0$ gauge
vector boson for the center-of-mass energy of $13$ $TeV$. The $PDF$
of Martin et al \cite{MMHT}, $MMHT2014-LO$, are used as the input
functions to feed the equations (\ref{eq7}). The results are the
double-scale $UPDF$ of the $KMR$ schemes. These $UPDF$ are in turn
substituted into the equation (\ref{eq4}) to construct the $Z$
cross-sections in the framework of $k_t$-factorization. One must
note that the experimental data of the $LHCb$ collaboration,
\cite{LHCb}, and the preliminary data of the $CMS$ collaboration,
\cite{CMS}, are produced in different dynamical setups; the $LHCb$
data are in the forward rapidity region, $2<|y_Z|<4.5$, while $CMS$
data are in a central rapidity sector, i.e. $0<|y_Z|<2.4$. We have
imposed the same restrictions in our calculations.

\begin{figure}[ht]
\centering
\includegraphics[scale=0.25]{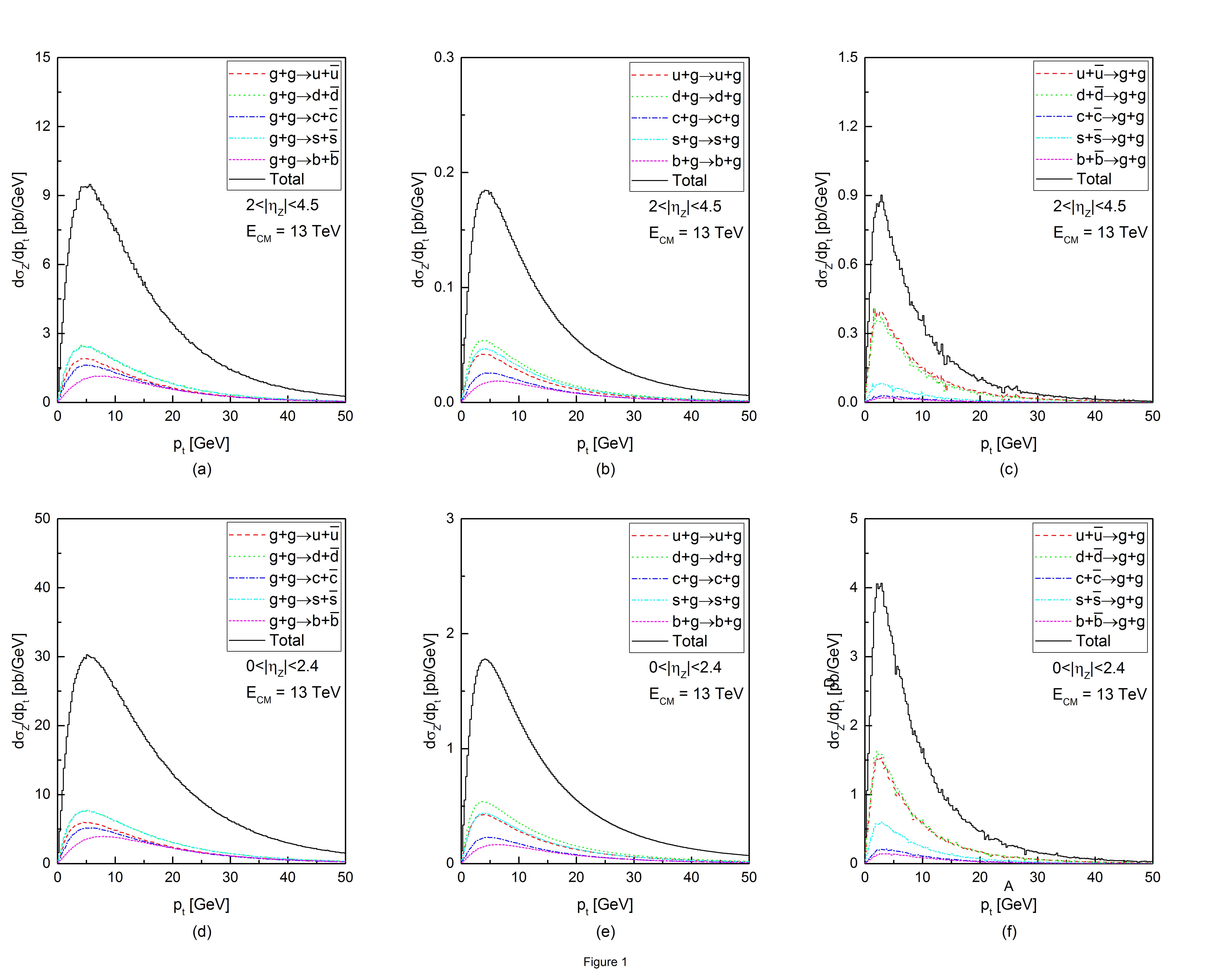}
\caption{Contributions of the individual quark flavors into the
differential cross-section of the productions of $Z^{0}$ boson in an
inelastic collision at $E_{CM}=13\;TeV$, plotted as a function of
the transverse momentum of the produced particle. The panels (a),
(b) and (c) illustrate our calculations for the forward rapidity
region, $2<|y_Z|<4.5$ (with the addition of $p_t^{\mu
\bar{\mu}}>20\;GeV$ and $60<m^{\mu \bar{\mu}}<120\;GeV$ constraints,
corresponding to the experimental measurements of the $LHCb$
collaboration, the reference \cite{LHCb}). The panels (d), (e) and
(f) are our results in the central rapidity region, $0<|y_Z|<2.4$
(with the addition of $p_t^{\mu \bar{\mu}}>25\;GeV$ and $60<m^{\mu
\bar{\mu}}<120\;GeV$ constraints, corresponding to the perlimanary
measurements of the $CMS$ collaboration, the reference \cite{CMS}).
The calculations are performed, using the $KMR$ $UPDF$ and the $PDF$
of $MMHT2014$.} \label{fig1}
\end{figure}

Thus, in the figure \ref{fig1} we present the reader with a
comparison between the different contributions into the differential
cross-sections of the production of $Z^0$, ($d\sigma_Z/dp_t$), as a
function of the transverse momentum ($p_t$) of the produced
particles, in the $KMR$ scheme. One readily notices that the
contributions from the $g^{*} + g^{*} \to Z^0 + q + \bar{q}$ (the
so-called gluon-gluon fusion process) dominate the the production.
The share of other production vertices is small (but not entirely
negligible) compared to these main contributions. This is to extent
different from our observations in the smaller center-of-mass
energies (see the section V of the reference \cite{NLO-W/Z}). Also,
differential cross-sections are considerably larger at the central
rapidity region compared to the results in the forward sector.

\begin{figure}[ht]
\centering
\includegraphics[scale=0.25]{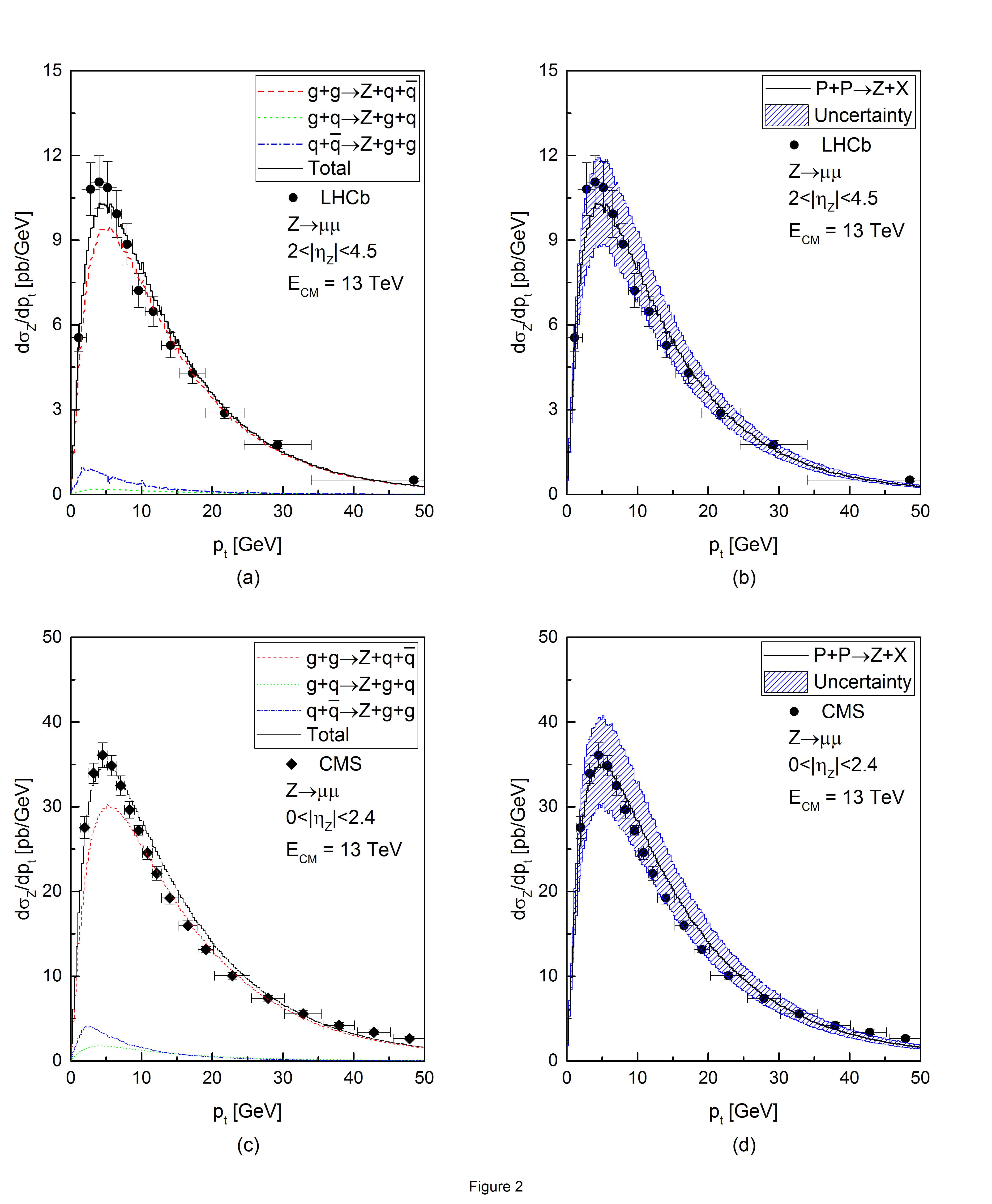}
\caption{Differential cross-section of the productions of $Z^{0}$
boson as a function of the transverse momentum of the produced boson
at $E_{CM} = 13\;TeV$. Panels (a) and (c) illustrate the
contributions from the individual sub-processes have been compared
to each other in the respective rapidity regions. The panels (b) and
(d) illustrate our results in their corresponding uncertainty
bounds, compared to the data of the $LHCb$ and the $CMS$
collaborations, the references \cite{LHCb,CMS}. The uncertainty
bounds have been calculated, by manipulating the hard-scale of the
$UPDF$ by a factor of 2.} \label{fig2}
\end{figure}

The total differential cross-section of the production of $Z^0$
vector boson is calculated within  the figure \ref{fig2}, as the sum
of the constituting partonic sub-processes (see the relation
(\ref{eq2})). The calculations are carried out for the
center-of-mass energy $E_{CM}=13\;TeV$ and plotted as a function of
the transverse momentum of the produced particle. In the panels (a)
and (c), the contributions from the individual sub-processes have
been compared to each other. The results in these panels
respectfully correspond to the forward rapidity region,
$2<|y_Z|<4.5$ (with the addition of $p_t^{\mu \bar{\mu}}>20\;GeV$
and $60<m^{\mu \bar{\mu}}<120\;GeV$ constraints, corresponding for
the experimental measurements of the $LHCb$ collaboration, the
reference \cite{LHCb}) and to the central rapidity region,
$0<|y_Z|<2.4$ (with the addition of $p_t^{\mu \bar{\mu}}>25\;GeV$
and $60<m^{\mu \bar{\mu}}<120\;GeV$ constraints, corresponding for
the perlimanary measurements of the $CMS$ collaboration, the
reference \cite{CMS}). The calculations have been performed, using
the $KMR$ $UPDF$ and the $PDF$ of $MMHT2014$. The panels (b) and (d)
illustrate our results in their corresponding uncertainty bounds,
compared to the data of the $LHCb$ and the $CMS$ collaborations. The
uncertainty bounds have been calculated, by means of manipulating
the hard-scale, $\mu$, of the $UPDF$ by a factor of 2, since this is
the only free parameter in our framework. Also, as expected for the
both regions, the contributions from the $g^{*}+g^{*} \to Z^0 + q +
\bar{q}$ sub-process dominate,
\begin{eqnarray}
    &&\hat{\sigma}(g^{*}+g^{*} \to Z^0 + q + \bar{q})
    \gg \hat{\sigma}(q^{*}+ \bar{q}^{*} \to Z^0 + g + g)
    >  \hat{\sigma}(g^{*}+q^{*} \to Z^0 + g +q).
\end{eqnarray}

\begin{figure}[ht]
\centering
\includegraphics[scale=0.25]{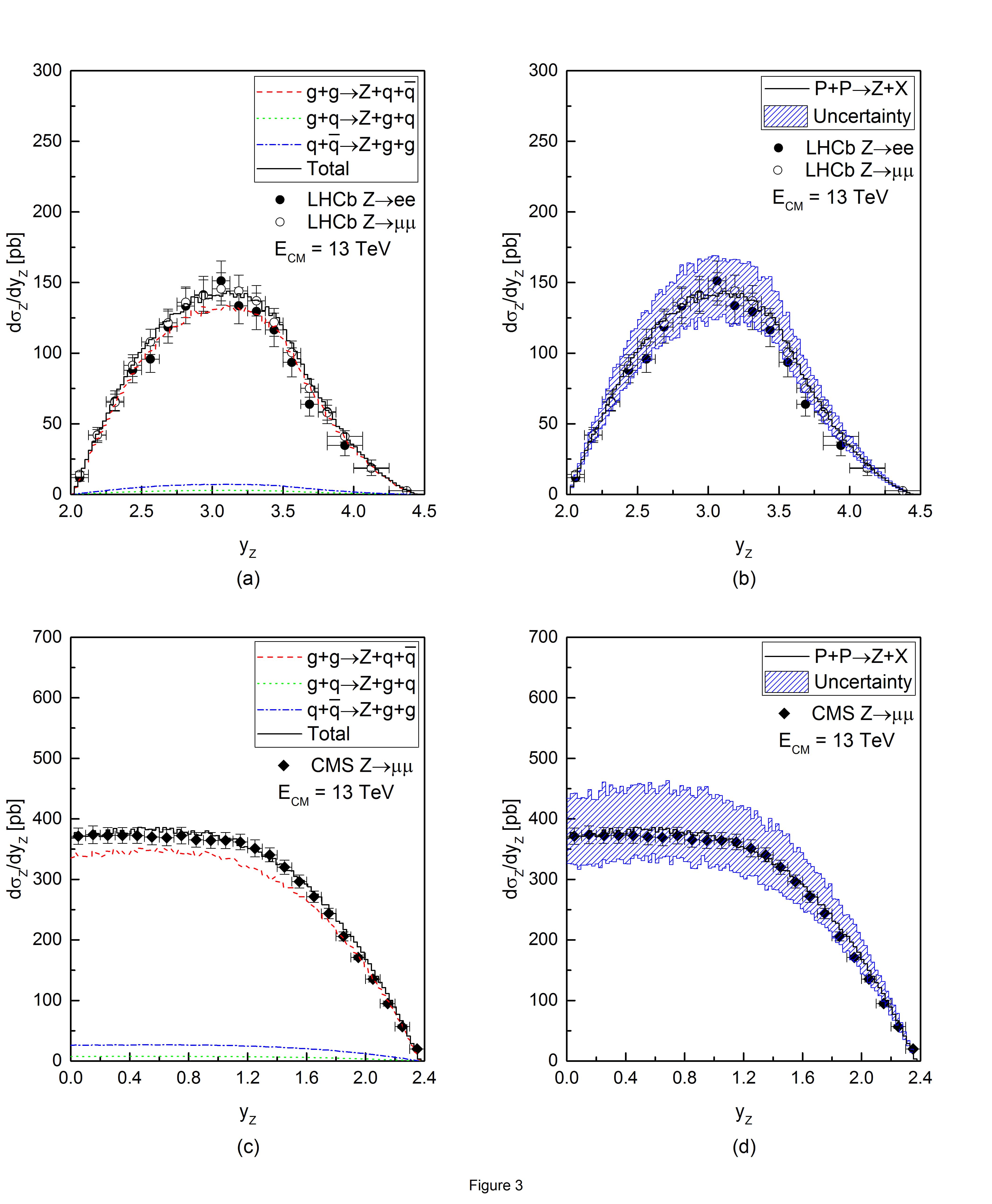}
\caption{Differential cross-section of the productions of $Z^{0}$
boson as a function of the rapidity of the produced boson at $E_{CM}
= 13\;TeV$. The notions of the diagrams are the same as in the
figure \ref{fig2}. } \label{fig3}
\end{figure}

The figure \ref{fig3} presents the differential cross-section of the
production of $Z^0$ vector boson, $d\sigma_Z/dy_Z$, as a function of
the rapidity of the produced boson ($y_Z$) at the center-of-mass
energy of $E_{CM}=13\;TeV$ in the $KMR$ formalism. The notion of the
figure is similar to that of the figure \ref{fig2}: The panels (a)
and (c) illustrate the contributions of each of the sub-processes
into the total production rate, while the total results have been
subjected to comparison with the experimental data of the $LHCb$ and
the $CMS$ collaborations (the references \cite{LHCb,CMS}), within
their corresponding uncertainty bounds, in the panels (b) and (d).
One finds that our calculations are in general agreement with the
experimental measurements.

Overall, it appears that our $NLO \otimes LO$ framework is generally
successful in describing the corresponding experimental measurements
in the explored energy range. This success if by part owed to the
$UPDF$ of $KMR$, which as an effective model, has been very
successful in producing a realistic theory in order to describe the
experiment, see the references \cite{NLO-W/Z, Modarres1, Modarres2,
Modarres3, Modarres4, Modarres5, Modarres6, Modarres7, Modarres8}.
One however should note that having a semi-successful prediction
from the framework of $k_t$-factorization by itself is a success,
since our calculations utilizing these $UPDF$ have inherently a
considerably larger error compared to those from the $NNLO$ $QCD$ or
even the $NLO$ $QCD$, presented here by the relatively large
uncertainty region. This is because we are incorporating the
single-scaled $PDF$ (with their already included uncertainties) to
form double-scaled $UPDF$ with additional approximations and further
uncertainties. Being able to provide predictions with a desirable
accuracy would require a thorough universal fit for these
frameworks, see the reference \cite{WattWZ}. Nevertheless, the
$k_t$-factorization framework, despite its simplicity and its
computational advantages, see the reference \cite{Modarres8,
WattWZ}, can provide us with a valuable insight regarding the
transverse momentum dependency of various high-energy $QCD$ events.

In summary, throughout the present work, we have calculated the
production rate of the $Z^0$ gauge vector boson in the framework of
$k_t$-factorization, using a $NLO \otimes LO$ framework and the
$UPDF$ of the $KMR$ formalism. The calculations have been compared
with the experimental data of the $LHCb$ and the $CMS$
collaborations. Our calculation, within its uncertainty bounds, are
in good agreement with the experimental measurements. We also
reconfirm that the $KMR$ prescription, despite its theoretical
disadvantages and its simplistic computational approach, has a
remarkable behavior toward describing the experiment.
\begin{acknowledgements}
$MM$ would like to acknowledge the Research Council of University of
Tehran and the Institute for Research and Planning in Higher
Education for the grants provided for him. $MRM$ sincerely thanks N.
Darvishi for valuable discussions and comments.
\end{acknowledgements}


\end{document}